\documentclass[a4paper]{article}

\usepackage{INTERSPEECH2020}
\usepackage{wrapfig}
\usepackage{subfigure}
\usepackage{multirow}
\usepackage{hhline}
\usepackage{todonotes}
\usepackage{floatrow}
\usepackage{float}

\title{Investigation of Large-Margin Softmax in Neural Language Modeling}
\name{Jingjing Huo$^1$, Yingbo Gao$^{1,2}$, Weiyue Wang$^{1,2}$, Ralf Schl{\"u}ter$^{1,2}$, Hermann Ney$^{1,2}$}

\address{$^1$Human Language Technology and Pattern Recognition Group, Computer Science Department \\
RWTH Aachen University, 52074 Aachen, Germany, $^2$AppTek GmbH, 52062 Aachen, Germany}

\email{\small \tt jingjing.huo@rwth-aachen.de, \{ygao, wwang, schlueter, ney\}@cs.rwth-aachen.de}

\begin{document}

\maketitle
\begin{abstract}
  To encourage intra-class compactness and inter-class separability among trainable feature vectors, large-margin softmax methods are developed and widely applied in the face recognition community. The introduction of the large-margin concept into the softmax is reported to have good properties such as enhanced discriminative power, less overfitting and well-defined geometric intuitions. Nowadays, language modeling is commonly approached with neural networks using softmax and cross entropy. In this work, we are curious to see if introducing large-margins to neural language models would improve the perplexity and consequently word error rate in automatic speech recognition. Specifically, we first implement and test various types of conventional margins following the previous works in face recognition. To address the distribution of natural language data, we then compare different strategies for word vector norm-scaling. After that, we apply the best norm-scaling setup in combination with various margins and conduct neural language models rescoring experiments in automatic speech recognition. We find that although perplexity is slightly deteriorated, neural language models with large-margin softmax can yield word error rate similar to that of the standard softmax baseline. Finally, expected margins are analyzed through visualization of word vectors, showing that the syntactic and semantic relationships are also preserved.
\end{abstract}

\section{Introduction}

The language model is an important component of automatic speech recognition (ASR) systems \cite{inproceedings_Kazuki, 6854535, Si2013PrefixTB, 7460091}, and perplexity (PPL) is known to be closely correlated with word error rate (WER) \cite{wilpon1994voice, Klakow:2002:TCW:638078.638080}.
Nowadays, state-of-the-art language models are commonly modeled using neural networks \cite{Bengio:2003:NPL:944919.944966, mikolov2010recurrent, Schwenk:2007:CSL:1230156.1230409, sundermeyer2015:lstm}.
The language model aims to learn the probability of word sequences, which are normally decomposed in an auto-regressive manner. To capture long contextual dependencies, the recurrent neural network (RNN) can be applied, which often uses the cross entropy training criterion along with softmax \cite{mikolov2010recurrent,sundermeyer2015:lstm, sundermeyer12:lstm}.

The idea of applying large-margin to the softmax layer is used to encourage intra-class compactness and inter-class separability among learned features. 
In the field of face recognition there exists a line of work  \cite{L-Softmax, Liang2017SoftMarginSF, CosFace, ArcFace, DBLP:journals/corr/abs-1801-05599, Liu_2019_CVPR} that studies large-margin in the softmax layer, providing significant improvements in performance. Considering that the vectors in the projection matrix before the last softmax layer in neural language models (NLM) are essentially feature vectors of the words, which resemble the feature vectors of images in face recognition, we are thus curious to examine the performances of the aforementioned margins in NLM.


Large-margin in NLM is not an unfamiliar concept. In \cite{LargeMarginNeuralLanguageModel}, a global-level margin that discriminates sentences is introduced. In contrast, this paper focuses on the margin between atomic-level word vectors. We apply different types of large-margins from face recognition to NLM. Our initial experiments show that using the largin-margin softmax from face recognition out-of-the-box for NLM deteriorates the PPL dramatically. We assume that this is due to the fundamental differences between words and faces in their class distributions. It is important to note that unlike in face recognition, the posterior probability of words in NLM is highly unbalanced. Zipf's law \cite{powers1998applications} is a common approximation of word frequency versus word rank in natural languages. In \cite{WeightNormInitialization}, the authors observe that NLM learn word vectors whose norms are closely related to word frequencies. Therefore, we conduct a series of experiments to compare various norm-scaling techniques for the word vectors. In addition, we implement a heuristic to scale the norms of the context vectors. It turns out that one of the norm-scaling methods slightly improves the PPL. When it is used along with the margin techniques, comparable WER to the baseline is achieved. Finally, to figure out the effects of margin techniques in NLM, we visualize the word vectors and observe that word vectors trained with large-margin softmax exhibit expected behaviors and ``stretch" the word vectors to more evenly populate the embedding space.

\section{Related Work}

The minimum distance of feature vectors to a decision boundary is called the margin. The large-margin concept plays an important role in conventional machine learning algorithms such as the support vector machine \cite{Support-vector-networks} and boosting algorithm \cite{schapire1998boosting}. Its core idea is to maximize the margin during training, in the hope that it leads to greater separability between classes during testing. \cite{chen2000discriminative, roark2004discriminative} study discriminative training on language models. The authors in \cite{sha2006large} introduce large margin into gaussian mixture models for phonetic classification and recognition task (multiway classification). Later in \cite{sha2007large}, they show a framework to train large margin hidden markov models in the more general setting of sequential (as opposed to multiway) classification in ASR. It has also been well studied in image processing. A novel loss function is proposed in \cite{NIPS2018_7364} to encourage large-margin in any set of layers of a deep network. In this work, we concentrate on the traditional margin methods that only focus on the output layer to see if it contributes to NLM.

The weights of the output layer are essentially feature vectors of each class (image features in face recognition or word embeddings in NLM). The scores (logits) of each sample are obtained using the dot product between the feature and the context vector. When using the cross entropy criterion along with softmax, the logits are used to calculate the loss. There exists a line of work in face recognition that modifies the loss function such that the scores of the true labels are reduced during training.

In \cite{L-Softmax}, the score for the ground truth class is manipulated by multiplying the angle between the ground truth feature vector and the context vector by a constant integer term $m$. It leads to a decline of that score, which ultimately leads to greater angular separability between learned feature vectors. This shares the similar idea with A-Softmax Loss (SphereFace) \cite{SphereFace}. However, SphereFace normalizes the weights by L2-norms in advance so that the learned features are restricted to be on a hypersphere manifold, which is consistent with the widely used assumption in the field of manifold learning for face recognition that face images lie on or close to a low-dimensional manifold \cite{Manifold_of_Facial_Expression, Face_recognition_using_Laplacianfaces}.

Later, Hao et al. \cite{CosFace} propose a large-margin cosine loss (CosFace) using L2 normalization for both the feature vectors and the context vector, and subtracting a margin $m$ from the cosine function output. CosFace also leads to a large-margin in the angular space. Subsequently, an additive angular margin loss (ArcFace) is presented in \cite{ArcFace} that adds a margin term $m$ to the angle instead of multiplying an integer term as in SphereFace. While these designs look similar, the authors claim that ArcFace has a better geometric attribute. Compared to SphereFace and CosFace, which are nonlinear in the angular space, ArcFace has a constant linear angular margin.

Word frequencies and vector norms are key concepts for this paper. Zipf's law \cite{powers1998applications} states that the frequency of words in a corpus of natural language is inversely related to the rank of words. Further in \cite{WeightNormInitialization}, the authors identify the relation between the norms of word vectors and their frequency. The result shows that the logarithm of word counts is a good approximation of word vector norms. This inspires us to examine various norm-scaling techniques for word vectors.

\section{Methodology}
\label{sec:methodology}
Assuming the long short-term memory (LSTM) network \cite{hochreiter1997long} is used for language modeling, the target word posterior probability using softmax in the last output layer can be written as:
\begin{equation}
P(y|i) = \frac{\exp{l(y,i)}}{\exp{l(y,i)} + \sum_{y' \neq y} \exp{l(y',i)}}
\end{equation}
where $y$ denotes the target next word (dependency on $i$ is dropped for simplicity), $y'$ is a running index in the vocabulary, $i$ is a running index of positions in data and $l(y,i)$ denotes the logit calculation. When using the inner-product, it can be written as:
\begin{align}
l(y, i) &= h_i^T W_{y} + b_{y}\\
        &= ||h_i|| \cdot ||W_{y}|| \cdot \cos \theta_{y,i} + b_{y}
\end{align}
where $h_i$ is the context vector and the output of the LSTM layer(s), $W_y$ is the embedding vector, $\theta_{y,i}$ is the angle between the two and $b_y$ is the bias term.

Commonly, the softmax output is used together with the cross entropy training criterion:
\begin{equation}
 L = - \sum_{i} \log P(y|i)
\end{equation}

\subsection{Conventional Margin}\label{sec:conventional_margin}

All of the three margins used in this paper only vary in the calculation of the logit of the ground-truth class $y$. For CosFace ($l_\text{COS}$) and ArcFace ($l_\text{ARC}$), the authors of the original paper claim that the normalization on features is necessary to encourage feature learning in their approach. Moreover, it is better to set the norm of the context vector as constant if the model is trained from scratch. Hence, they set $||W_y|| = 1$ and $||h_i|| = s$, where $s$ is some predefined constant. In contrast to ArcFace and CosFace, L-Softmax ($l_\text{LSM}$) does not normalize anything in advance. The three margins from face recognition are formally defined as follows:
\begin{align} 
l_{\text{COS}}(y, i) &= s \times (\cos(\theta_{y,i}) - m) \label{CosFace_fun}\\
l_{\text{ARC}}(y, i) &= s \times \cos(\theta_{y,i} + m) \label{ArcFace_fun}\\
l_{\text{LSM}}(y, i) &=  ||h_i|| \cdot ||W_{y}|| \cdot \varphi(\theta_{y,i})
\label{L-Softmax_fun}
\end{align}
where $\varphi(\theta_{y,i})$ is designed as:
\begin{equation}
    \varphi(\theta_{y,i}) = (-1)^k \cos(m \theta_{y,i}) - 2k
\label{varphi}
\end{equation}
 While margin $m$ in $l_{\text{COS}}$ and $l_{\text{ARC}}$ are non-negative real numbers, it must be a positive integer in $l_{\text{LSM}}$. Using (\ref{varphi}), the monotonicity of $\varphi(\theta_{y,i})$ with respect to $m$ can be guaranteed. $k$ in $\varphi(\theta_{y,i})$ is an integer in the range of $[0,m-1]$ and $\theta \in [\frac{k\pi}{m}, \frac{(k+1)\pi}{m}]$.

\subsection{Margin with Norm-scaling}
We explore different norm-scaling techniques for word vectors and context vectors, which differ only in how they alter the norms of vectors. $f(y)$ defines the modifications to word vectors and $g(i)$ defines the modifications to context vectors:
\begin{align}
    f(y) &= 
    \begin{cases}
    ||W_y||, & \textit{no-mod}\\
    ||W_{\text{argmax}\{c\}}||, & \textit{uniform} \\
    \log (\mathrm{exp}(||W_{\text{argmax}\{c\}}||) - v \times y), & \textit{log-rank} \\
    ||W_{\text{argmin}\{c\}}|| + u \times c_y,  & \textit{unigram}\\
    \log (c_y),  & \textit{log-unigram} 
    \label{norm_scaling_word_vector}
    \end{cases} \\
    g(i) &= 
    \begin{cases}
    ||h_i||, & \textit{no-mod}\\
    \text{max}\{||h||\}, & \textit{max-norm} 
    \end{cases}
\end{align}
\noindent with $v$ and $u$ defined as:
\begin{align} 
v &= \frac{\mathrm{exp}(||W_{\text{argmax}\{c\}}||) - \mathrm{exp}(||W_{\text{argmin}\{c\}}||)}{V} \\
u &= \frac{||W_{\text{argmax}\{c\}}|| - ||W_{\text{argmin}\{c\}}||}{\text{argmax}\{c\}}
\end{align}
\noindent where $c$ is the count of words and $V$ is the vocabulary size.

For \textit{uniform} we assume that all word vectors have the same norm, and in this case we use the norm of the word vector with the largest count among all words in the training corpus. For \textit{log-rank}, we expect that the norm of the word vector is linear with respect to its index before a logarithmic operation (assuming words are sorted in descending order by their counts). For \textit{unigram}, we use scaled and shifted word counts as new word vector norms after normalization. Note that for \textit{uniform}, \textit{log-rank} and \textit{log-unigram}, $f(y)$ is dynamically updated in each update step during training. For \textit{log-unigram}, we take the logarithm of word count directly as the norm of that word. On the other hand, the heuristic \textit{max-norm} scales the norm of the $i$-th context vector $h_i$ using the largest norm in the batch where the $h_i$ appears.

Finally, combining norm-scaling and margin techniques, the logit calculation can be reformulated as:
\begin{equation}
     l(y, i) =  g(i) f(y) \phi (\theta_{y, i})  + b_y
\end{equation}
where $g(i)$ is selected between $||h_i||$ and $\max\{||h||\}$ , $f(y)$ is selected among the five norm-scaling functions and $\phi (\theta_{y', i}) =  \cos (\theta_{y', i})$ for $y' \neq y$, otherwise
\begin{equation}
    \phi (\theta_{y, i}) = 
    \begin{cases}
    \cos(\theta_{y,i}) - m, & \text{COS} \\
    \cos(\theta_{y,i} + m), & \text{ARC} \\
    \varphi(\theta_{y,i}),  & \text{LSM}\\
    \cos (\theta_{y, i}), & \textit{no-margin}
    \end{cases}
\end{equation}
For all experiments in the next section, we always keep the bias term $b_y$, as according to our early experiments, dropping it slightly degradates the performance across all setups.

\section{Experiments}
\label{sec:experiments}
We use two datasets to compare the effects of the aforementioned techniques: Switchboard (SWB) and Quaero English. SWB is a relatively small dataset, with a vocabulary size of 30K and 25M training tokens. Quaero has a vocabulary size of 128K and 49M training tokens. We use two-layer LSTM language models with hidden state sizes of 1024 and 2048 for SWB and Quaero, respectively. For Quaero, we also apply the sampled softmax \cite{DBLP:journals/corr/JeanCMB14} method to speed up training. In the following experiments, we fix the model architecture and only alter the softmax layer.

\subsection{Conventional Margin}

Considering that large-margin works well in face recognition, to grasp the preliminary understanding of its effects on our task we apply the large-margin techniques described in Section \ref{sec:conventional_margin} out-of-the-box for NLM. As shown in Table \ref{PPLSWBLARGIN_MARGIN}, for \text{LSM}, the norms of vectors are retained as defined in Equation \ref{L-Softmax_fun}, and setting $m$ as one means that there is no modification of cosine similarity. Therefore, the first row of the table gives us the baseline of this margin. Moreover, as required in \cite{L-Softmax} that $m$ must be an integer, the minimal step is to increase $m$ by one. As can be seen, even setting $m$ to two would dramatically worsen the performance, we do not further increase $m$ in this experiment.

\begin{table}[ht]
\centering
\caption{PPL on SWB using margins from face recognition out-of-the-box.}
\begin{tabular}{|l|r|r|}
\hline
Method &  $m$             & PPL \\ \hline
Baseline & n/a  & 53.7   \\ \hline
\multirow{2}{*}{LSM} & 1     & 53.5 \\\cline{2-3} 
                           & 2     & 390.3                    \\ \hline
\multirow{5}{*}{ARC}   &  0                & 66.9   \\ \cline{2-3} 
                            & 0.001 & 68.0              \\\cline{2-3} 
                           & 0.003 &  80.7              \\\cline{2-3} 
                           & 0.01  & 106.1             \\\cline{2-3} 
                           &  0.03  & 170.4              \\ \hline
\multirow{5}{*}{COS}   &  0                & 66.9   \\\cline{2-3} 
                            & 0.001 & 70.3             \\\cline{2-3} 
                           & 0.003 & 77.7              \\\cline{2-3} 
                           & 0.01  & 108.0             \\\cline{2-3} 
                           & 0.03  & 382.1             \\ \hline
\end{tabular}
\label{PPLSWBLARGIN_MARGIN}
\end{table}
For COS and ARC, the modifications on feature vector norms are defined in Equation \ref{CosFace_fun} and Equation \ref{ArcFace_fun}. The feature vector $W_y$ and context vector $h_i$ are firstly normalized and then a large scalar $s=64$ is used for re-scaling. The results show a clear trend that the bigger the margin term gets, the worse the performance is. Even when disabling the margin, i.e. $m = 0$, PPL is much higher than the baseline. As the only changes to the calculation in this case is the re-scaling of vector norms, this suggests that normalizing the word vectors and re-scaling them to have norms of 64 is too harsh for NLM in concern. 

\subsection{Margin with Norm-scaling}
Considering the pattern discovered in \cite{WeightNormInitialization}, that the norms of word vectors approximate the logarithm of word counts, as well as the results of our preliminary experiments, we believe that norms of vectors play a nonnegligible role in NLM. Simply reducing them and re-scaling them using a large constant seems improper in our case. Hence, our next step aims to figure out which kind of norm-scaling is more suitable for NLM. Specifically, we vary the norm-scaling setup on $g(i)$ and $f(y)$ and examine the PPL of the corresponding models.

The top half of Table \ref{norm-scaling} depicts the performance of five different norm-scaling techniques of word vectors as defined in (\ref{norm_scaling_word_vector}), where $g(i)$ uses \textit{no-mod}.
As seen, all of them slightly worsen the performance. 
The bottom half of the table reports the performance when $g(i)$ uses \textit{max-norm}. We can see that only applying the heuristic on the context vector gives the best performance on SWB, and using \textit{max-norm} and \textit{log-unigram} together slightly improves the PPL on Quaero. We go on to apply the best norm-scaling setup in combination with  $\phi (\theta_{y, i})$ variants for NLM. 

\begin{table}[ht]
\centering
\caption{PPL on SWB or Quaero using different norm-scaling techniques. $g(i)$ and $f(y)$ being \textit{no-mod} corresponds to the standard softmax baseline.}
\begin{tabular}{|l|l|cc|}
\hline
\multicolumn{1}{|c|}{$g(i)$}                         & \multicolumn{1}{|c|}{$f(y)$}           & \multicolumn{1}{c}{SWB} & \multicolumn{1}{c|}{Quaero} \\ \hline
\multirow{5}{*}{\textit{no-mod}}   & \textit{no-mod}      & 53.7                    & 105.8                    \\\cline{2-4} 
                          & \textit{uniform}     & 56.8                    & 108.4                    \\\cline{2-4} 
                          & \textit{log-rank}  & 56.8                    & 108.1                    \\\cline{2-4} 
                          & \textit{unigram}     & 56.0                    & 108.4                    \\\cline{2-4} 
                          & \textit{log-unigram} & 53.8                    & 107.4                    \\ \hline
\multirow{5}{*}{\textit{max-norm}} & \textit{no-mod}      & \textbf{52.9}                    & 104.3                    \\\cline{2-4} 
                          & \textit{uniform}     & 56.4                    & 107.2                    \\\cline{2-4} 
                          & \textit{log-rank}  & 57.4                    & 109.6                    \\\cline{2-4} 
                          & \textit{unigram}     & 54.6                    & 106.1                    \\\cline{2-4} 
                          & \textit{log-unigram} & 53.1                    & \textbf{104.1}                    \\ \hline
\end{tabular}
\label{norm-scaling}
\end{table}

\begin{table}[ht]
\centering
\caption{PPL on SWB combining norm-scaling and large-margin softmax.}
\begin{tabular}{|l|l|l|cc|}
\hline
\multicolumn{1}{|c|}{$g(i)$}      & \multicolumn{1}{|c|}{$f(y)$}  & \multicolumn{1}{|c|}{$m$}    & ARC       & COS       \\ \hline
\multirow{2}{*}{\textit{no-mod}}   & \textit{no-mod}                  & 0.001 & 54.5          & 54.5          \\ \cline{2-5} 
                          & \textit{log-unigram}             & 0.001 & 55.3          & 55.3          \\ \hline
\multirow{5}{*}{\textit{max-norm}} & \textit{log-unigram}             & 0.001 & 55.3          & 54.9          \\ \cline{2-5} 
                          & \multirow{4}{*}{\textit{no-mod}} & 0.001 & \textbf{54.2} & \textbf{54.1} \\ \cline{3-5} 
                          &                         & 0.003 & 55.5          & 55.9          \\ \cline{3-5} 
                          &                         & 0.006 & 57.7          & 58.4          \\ \cline{3-5} 
                          &                         & 0.010  & 60.0          & 60.9          \\ \hline
                     
\end{tabular}
\label{ns_with_lm_SWB}
\end{table}

\begin{table}[ht]
\centering
\caption{PPL on Quaero combining norm-scaling and large-margin softmax.}
\begin{tabular}{|l|l|l|c|c|}
\hline
\multicolumn{1}{|c|}{$g(i)$}                         & \multicolumn{1}{|c|}{$f(y)$}     & \multicolumn{1}{|c|}{$m$}      & ARC        & COS        \\ \hline
\multirow{2}{*}{\textit{no-mod}}   & \textit{no-mod}   & 0.001  & \textbf{111.2}          & \textbf{111.6}          \\ \cline{2-5} 
                          & \textit{log-unigram}  & 0.001  & 114.7          & 114.2          \\ \hline
\multirow{2}{*}{\textit{max-norm}} & \textit{log-unigram}   & 0.001 & 113.7 & 112.7          \\ \cline{2-5} 
                          & \textit{no-mod}    & 0.001     & 114.0          & 112.0 \\ \hline
\end{tabular}
\label{ns_with_lm_Quaero}
\end{table}

Now that we have good norm-scaling setups for both the context vectors and the word vectors, the logical next step is to assess the performance of various margins in combination with them. We choose the four best combinations of $g(i)$ and $f(y)$ in Table \ref{norm-scaling}, and conduct large-margin experiments. First, we use a very small margin term for all of them, i.e. $m=0.001$. As can be seen in the first four rows in Table \ref{ns_with_lm_SWB}, they do not differ much in PPL and none of them improves over the baseline on SWB. Furthermore, as shown in Table \ref{ns_with_lm_Quaero}, all of them deteriorate the PPL on Quaero to a large degree. To further verify, we tune the margin term under our best norm-scaling setting. The results in the last five rows in the Table \ref{ns_with_lm_SWB} clearly show that the performance gets worse as $m$ increases.

Last but not the least, we conduct LSTM recurrent neural network rescoring experiments as shown in Table \ref{asr} to make a final verdict of the application of large-margin softmax in NLM. The baseline system is based on the hybrid hidden Markov model neural network\cite{kitza19:interspeech}. It is interesting to find that although PPL is deteriorated, NLM with large-margin softmax can yield the same WER as the baseline.

\begin{table}[ht]
\centering
\caption{PPL and WER on SWB using ARC or COS with $m = 0.001$ and best norm-scaling techniques.}
\begin{tabular}{|l|c|r|r|r|}
\hline
\multicolumn{2}{|c|}{Metrics} & baseline & ARC & COS \\ \hhline{|==|=|=|=|}
\multicolumn{2}{|l|}{PPL}     & 52.9      & 54.2    & 54.1    \\ \hhline{|==|=|=|=|}
\multirow{3}{*}{WER}   & Switchboard   & 13.7      & 13.7     & 13.7      \\ \cline{2-5} 
                       & Callhome   & 7.1     &  7.1       & 7.1        \\ \cline{2-5} 
                       & Average  & 10.4    &  10.4       &  10.4       \\ \hline
\end{tabular}
\label{asr}
\end{table}


\section{Analysis}
To analyze the effects of large-margin in NLM, in this section we visualize the word embeddings trained with large-margin softmax as well as the standard softmax. For visualization, the dimensionality of word vectors is reduced to two by first applying principal component analysis and then using t-distributed stochastic neighbor embedding \cite{maaten2008visualizing}.

\begin{figure}[ht]
\centering
\subfigure{
\begin{minipage}[t]{0.39\linewidth}
\centering
\includegraphics[width=\textwidth]{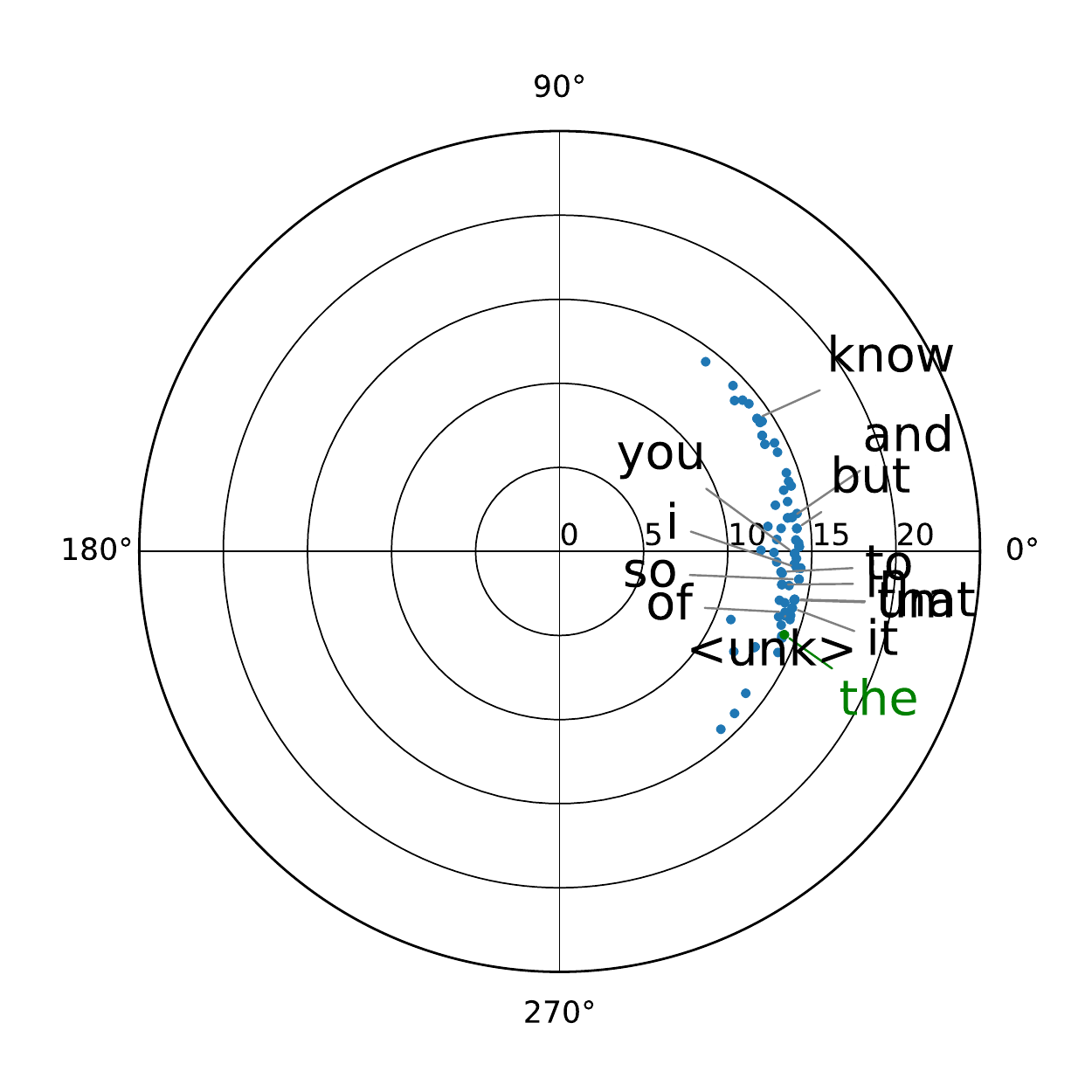}
\centerline{(a) softmax, top 100}
\end{minipage}
}
\subfigure{
\begin{minipage}[t]{0.39\linewidth}
\centering
\includegraphics[width=\textwidth]{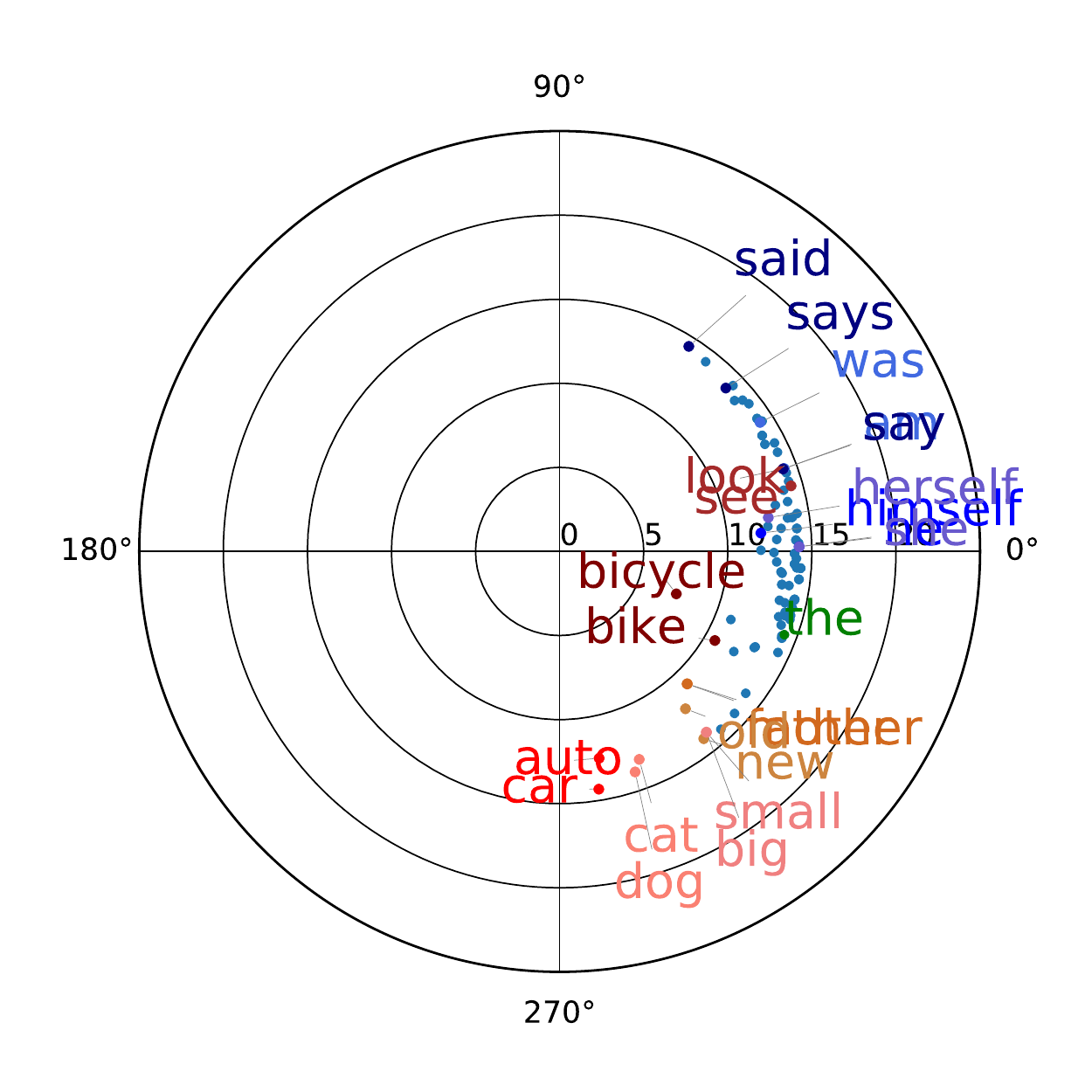}
\centerline{(d) softmax, word groups}
\end{minipage}
}
\subfigure{
\begin{minipage}[t]{0.39\linewidth}
\centering
\includegraphics[width=\textwidth]{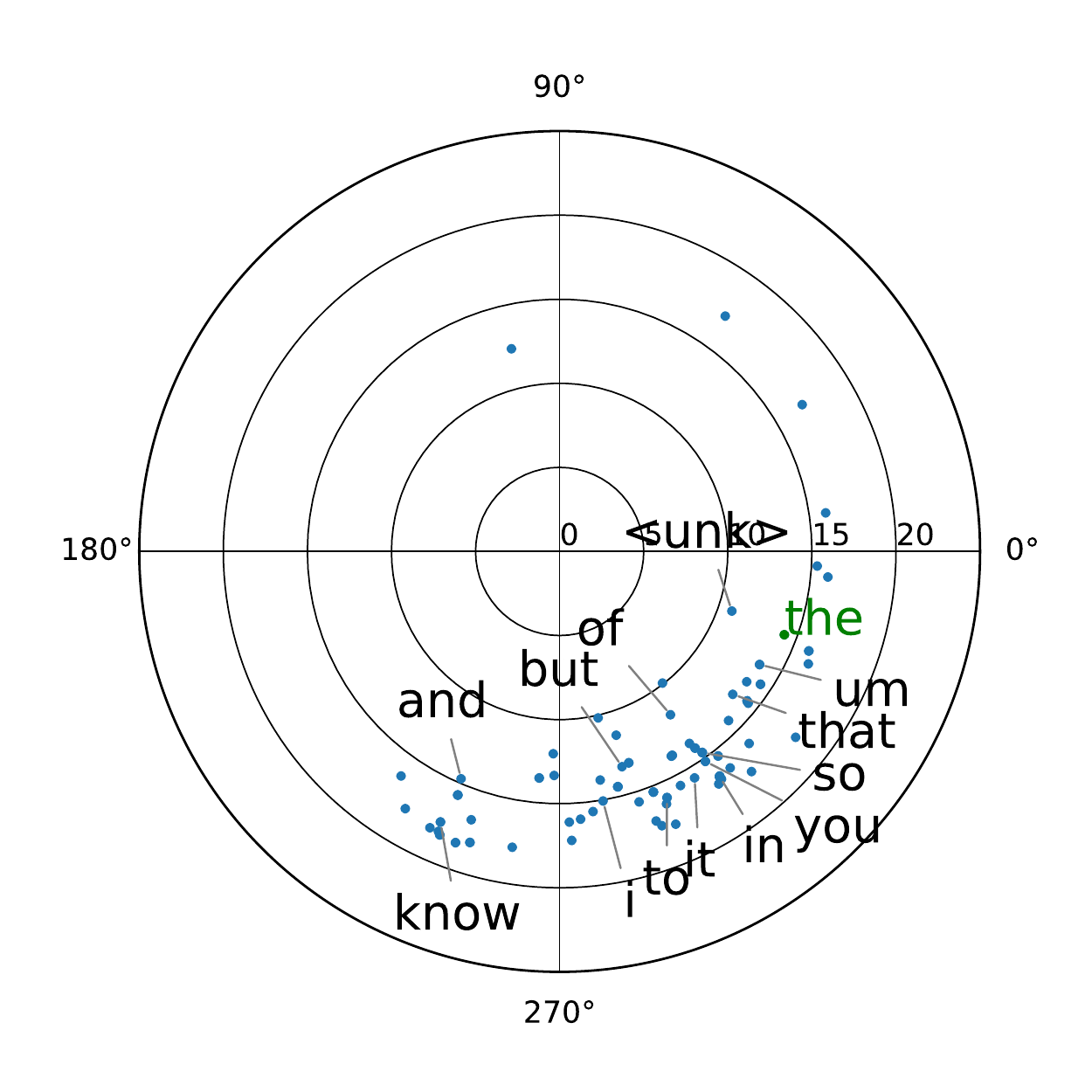}
\centerline{(b) COS, top 100}
\end{minipage}
}
\subfigure{
\begin{minipage}[t]{0.39\linewidth}
\centering
\includegraphics[width=\textwidth]{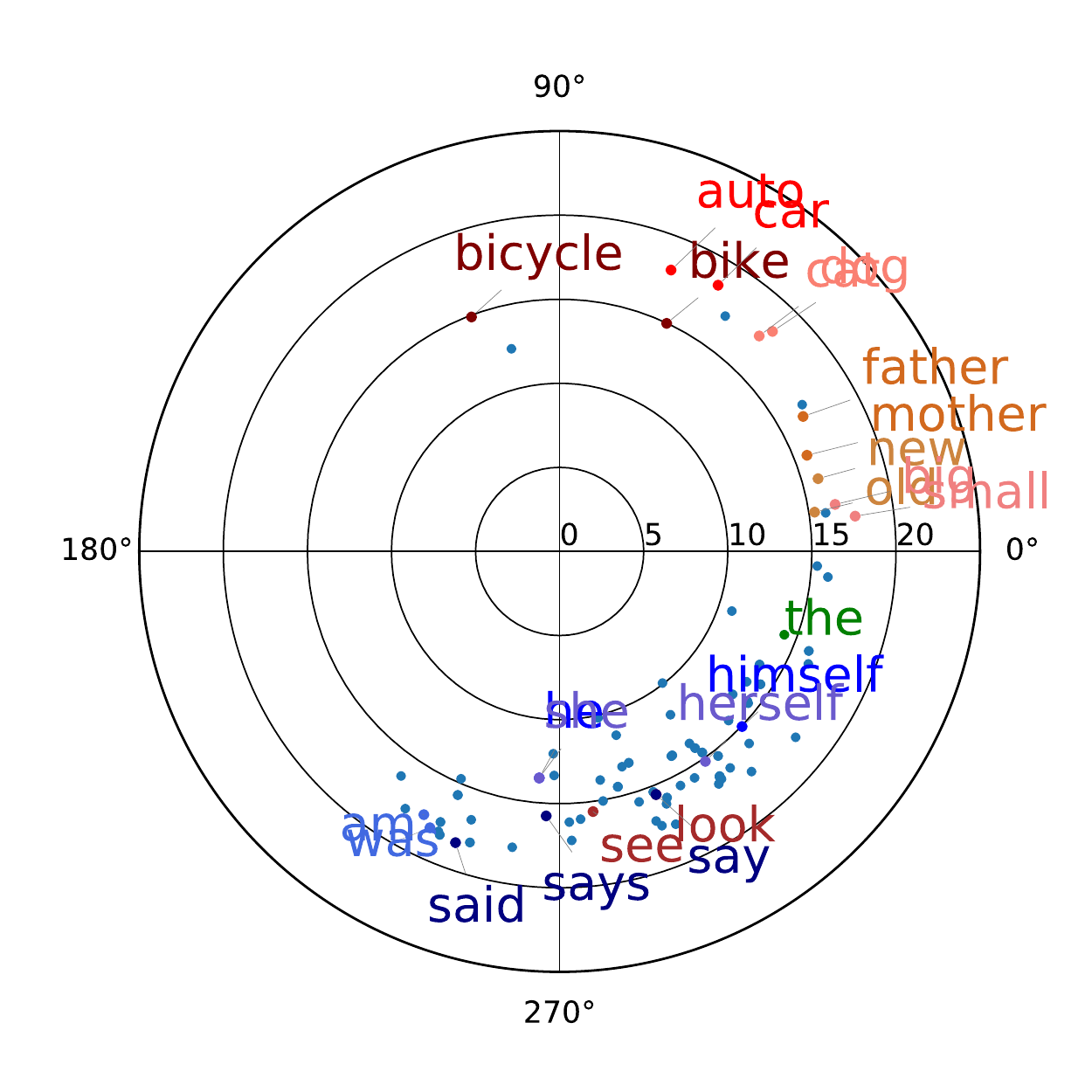}
\centerline{(e) COS, word groups}
\end{minipage}
}
\subfigure{
\begin{minipage}[t]{0.39\linewidth}
\centering
\includegraphics[width=\textwidth]{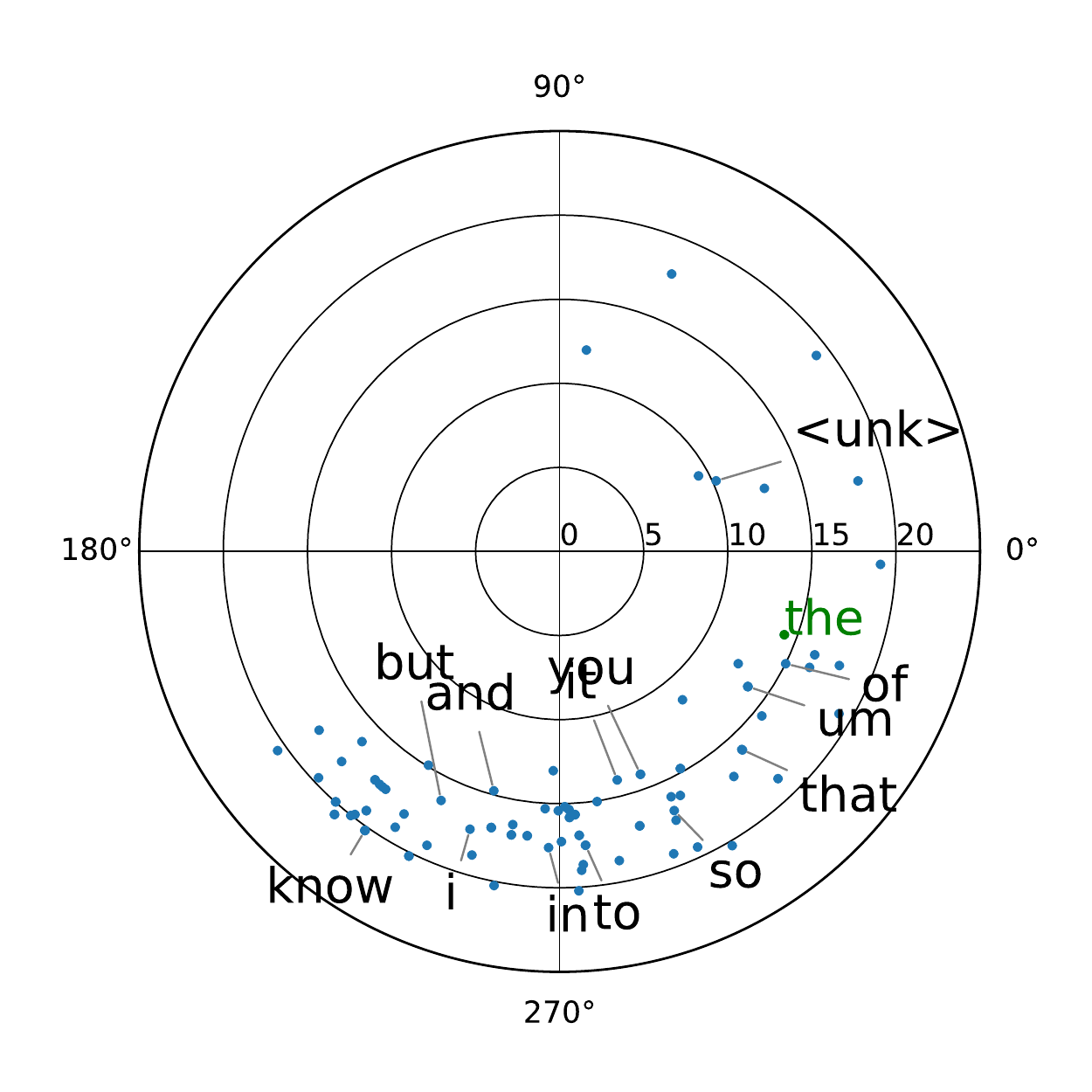}
\centerline{(c) ARC, top 100}
\end{minipage}
}
\subfigure{
\begin{minipage}[t]{0.39\linewidth}
\centering
\includegraphics[width=\textwidth]{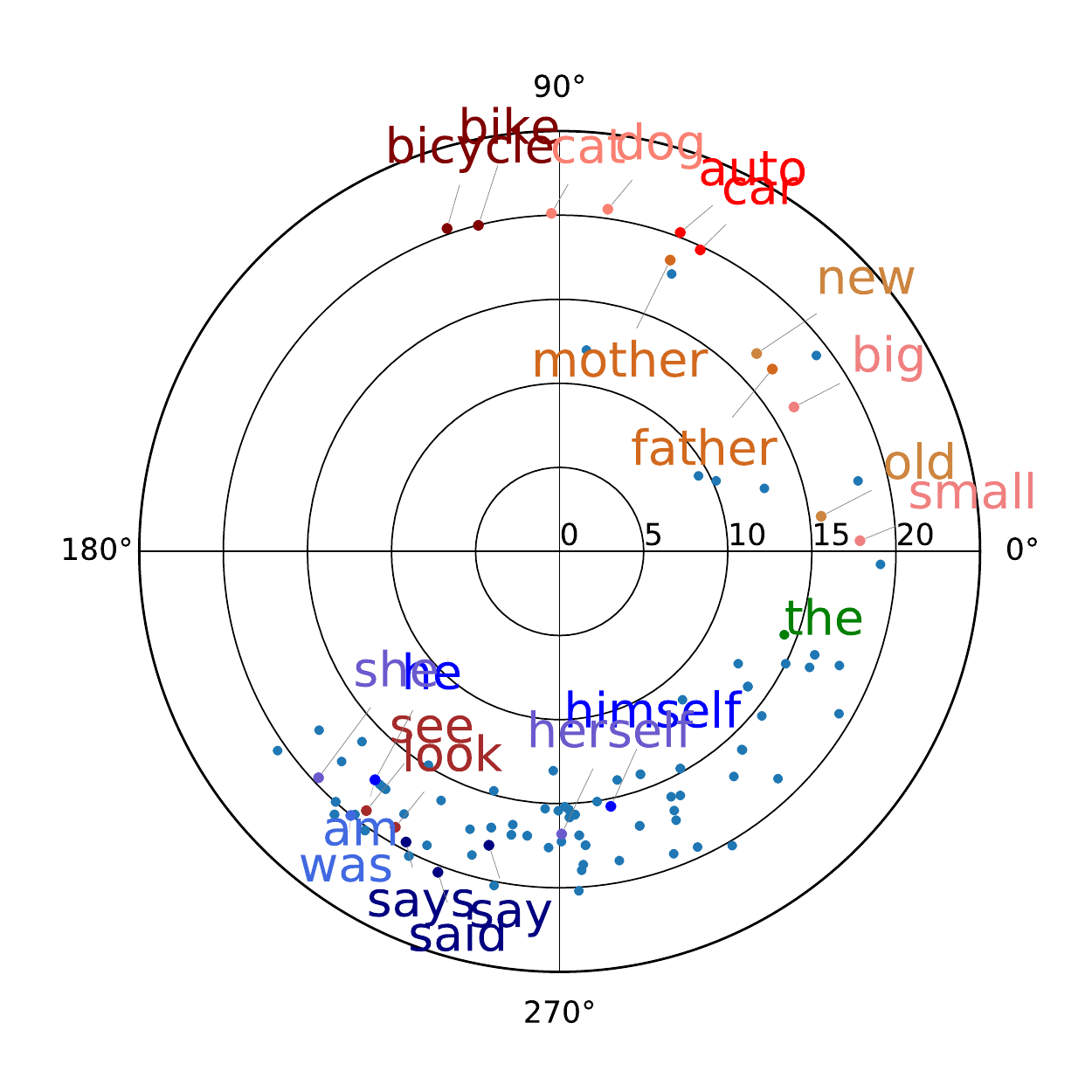}
\centerline{(f) ARC, word groups}
\end{minipage}
}
\centering
\caption{Word vectors plotted in polar coordinates. In (a), (b) and (c), the top 100 frequent words more evenly populate the embedding space. In (d), (e) and (f), word groups with strong semantic (shades of red) or syntactic (shades of blue) relations are preserved.}
\label{fig:word_vector}
\end{figure}

Figure \ref{fig:word_vector} shows the word vectors in polar coordinates. For COS and ARC, we use the large-margin softmax in bold in Table \ref{ns_with_lm_SWB}. The vectors are scaled and rotated to align the word ``the" in all plots. The points in blue are the top 100 frequent words in SWB, which already account for around 65\% of the total running words.  As can be seen, these vectors of frequent words obtained by large-margin softmax approaches in (b) and (c) are more separable than those obtained by standard softmax in (a). In other words, the word vectors are further ``stretched" to more evenly populate the embedding space.

To further investigate if the word embeddings obtained by large-margin softmax maintain the word relations in general. We visualize some word groups in the second column ((d), (e) and (f)) of Figure \ref{fig:word_vector}. The words in red color series are pairs that have semantic similarity while word groups in blue color series have syntactic relations. As seen, even though the large-margin softmax makes the angles among words larger, it can still preserve the semantic and syntactic relations. For instance, words that share a similar meaning (``auto" - ``car") are well gathered and the angle between word ``she" and ``herself' is almost the same as the angle between word ``he" and ``himself" in (f). 

\section{Conclusions}

In this work, we investigate the use of large-margin softmax in neural language models. We first apply margins from face recognition out-of-the-box, which evidently deteriorates perplexity. Considering the unbalanced nature of word distributions, we further conduct experiments to find good norm-scaling settings for neural language models and tune the margin parameters. Then we apply the models trained with large-margin softmax in rescoring experiments, where we can reach the same word error rate performance as the standard softmax baseline. Finally, to figure out the effects of large-margin in neural language models, we visualize the word vectors. It is interesting to note that the expected margin are found among the word vectors trained with large-margin softmax, which makes them more evenly populate the embedding space. At the same time, the semantic and syntactic relations among words are also preserved.

\section{Acknowledgements}


{\footnotesize This project has received funding from the European Research Council (ERC) under the European Union’s Horizon 2020 research and innovation programme (grant agreement No 694537, project ``SEQCLAS"). The work reflects only the authors' views and the European Research Council Executive Agency (ERCEA) is not responsible for any use of that may be made of the information it contains.}

\bibliographystyle{IEEEtran}
\bibliography{template}

\end{document}